\documentclass[12pt]{amsart}
\usepackage{graphicx}
\usepackage{geometry}
\usepackage{amssymb}
\usepackage{setspace}
\usepackage[dvipsnames]{pstricks}
\usepackage{pst-node}
\usepackage{pst-slpe}
\usepackage{pst-tree}
\usepackage{bbm}
\usepackage{amsmath, amsthm, amssymb}
\usepackage{pgfplots}
    \pgfplotsset{width=8cm,compat=1.5.1}
\usetikzlibrary{arrows.meta}
    \usetikzlibrary{shapes,snakes}
\usepackage{amsmath,amssymb,amsthm}
\usepackage{tikz}
\usepackage{caption}
\usepackage{subcaption}
\usepackage{booktabs}
\usepackage{empheq}
\usepackage[most]{tcolorbox}
\usepackage{diagbox}

\newtcbox{\mymath}[1][]{%
    nobeforeafter, math upper, tcbox raise base,
    enhanced, colframe=blue,
    colback=yellow!30, boxrule=1pt,
    #1}
    
\usepackage{natbib}
\usepackage{draftwatermark}
\SetWatermarkText{}
\SetWatermarkScale{4}

\newtheorem{theorem}{Theorem}

\newtheorem{axiom}{Axiom}

\newtheorem{claim}{Claim}

\newtheorem{example}{Example}

\newtheorem{lemma}{Lemma}

\newtheorem{proposition}{Proposition}
\newtheorem{remark}{Remark}

\newenvironment{definition}[1][Definition]{\noindent\textbf{#1.} }{}
\usepackage{tikz}
\renewcommand{\emptyset}{\varnothing}

\newcommand{\call}{\mathcal{L}}
\begin{document}
\title{Correlated Choice}
\author{Christopher P. Chambers, Yusufcan Masatlioglu, \\ Christopher Turansick}
\thanks{We thank the coeditor, Todd Sarver, as well as two anonymous referees for their comments which have helped improve this paper. We also thank Tugce Cuhadaroglu, Federico Echenique, Ricky Li, Jay Lu, Paola Manzini, John Rust, Tomasz Strzalecki, Francis Vella, and seminar participants at RUD 2021 and SAET 2022 for helpful comments and discussions. We are especially grateful to Peter Caradonna for pointing out an error in an earlier version.\\
Chambers:  Department of Economics, Georgetown University, ICC 580  37th and O Streets NW, Washington DC 20057. E-mail: \texttt{Christopher.Chambers@georgetown.edu}.
Masatlioglu:  University of Maryland, 3147E Tydings Hall, 7343 Preinkert Dr.,  College Park, MD 20742. E-mail: \texttt{yusufcan@umd.edu}
Turansick:  Department of Economics, Georgetown University, ICC 580  37th and O Streets NW, Washington DC 20057.  E-mail:  \texttt{cmt152@georgetown.edu} \\
A prior version of this paper has been circulated under the name ``Influence and Correlated Choice." }
\date{\today}
\maketitle
\begin{abstract}
    We study random joint choice rules, allowing for interdependence of choice across agents.  These capture random choice by multiple agents, or a single agent across goods or time periods. %Of interest are random joint choice rules which have well defined marginal random choice rules.% as this is necessary for choices to be separable over each dimension. 
    Our interest is in separable choice rules, where each agent can be thought of as acting independently of the other.  
    A random joint choice rule satisfies marginality if for every individual choice set, we can determine the individual's choice probabilities over alternatives independently of the other individual's choice set. We offer two characterizations of random joint choice rules satisfying marginality in terms of separable choice rules. While marginality is a necessary condition for separability, we show that it fails to be sufficient. We provide an additional condition on the marginal choice rules which, along with marginality, is sufficient for separability.
\end{abstract}

\section{Introduction}
To researchers and analysts, choice data often appear stochastic. An explanation for this is that choices are in fact deterministic conditional on some latent component to preference.  The latent variable is randomly determined, observed by the decision maker, and unobserved by the economist. The distribution of this variable then induces a distribution over choices. For example, choice data over umbrellas may appear stochastic if weather data is unobserved.\footnote{While our main motivation is this latent variable story, other explanations of stochastic choice are fluctuating tastes (\cite{thurstone1927law, luce1959individual, block1959random}),  random bounded rationality (\cite{Filiz-Ozbay2020ProgressiveChoice}), just-noticeable-difference (\cite{horan2021stochastic}), random attention (\cite{manzini2014stochastic, Cattaneo-Ma-Masatlioglu-Suleymanov_2020_JPE}), learning (\cite{baldassi2020behavioral}), random stopping (\cite{dutta2020gradual}), imperfect information (\cite{natenzon2019}), random attributes (\cite{Gul_Natenzon_Pesendorfer_2014_ECMA}), random reference point (\cite{Kibris2022}), or deliberate randomization (\cite{machina1985stochastic, cerreia2019deliberately,allen2021revealed}).}

%In this sense, the idea of random utility maximization can be interpreted as follows. 

%here is an underlying distribution over unobservables and, conditional on a realization, the decision maker maximizes their utility function, which depends on the unobservable.  

Consistent with this observation is the idea that multiple decision makers may condition their decision on the same unobserved component.  Examples abound:  if a seasonal component to data is unobserved, we would expect two individuals living in the same hemisphere to tend to choose, from a pair consisting of snow boots or flip flops, the flip flops roughly in the same observations.  An individual living in the northern hemisphere would choose the flip flops when the individual in the southern hemisphere chooses the snow boots, and vice versa.  If location data is unobserved, two people in the same neighborhood would prefer to locate a dump in another neighborhood, and an opera house in their neighborhood.  If they lived in opposite neighborhoods, the choices would be opposite.  All of these examples have the feature that the unobserved component leads choices to be apparently random to an analyst, but there is some observable structure to the choices across the agents:  an observation tells us what each of the agents has chosen from their corresponding choice sets.  This observation leads us to enrich the classical choice model, allowing choice data to be joint. Our main goal will be to study a stochastic version of this model.%in a stochastic form.

%If we are to believe this story, it should be the case that, when we add a second decision maker, the choices of the two decision makers are governed by the same underlying distribution over unobservables.

%We study a special case of this story. 

A degenerate joint choice function specifies, for each pair of choice sets a pair of agents may face, a unique choice for each of them.  Joint choice functions are flexible enough to allow the agents to interact, possibly allowing their choices to depend on the other agent's choice set.  However, our interest is in the simple setting where each individual has their own choice function over their own choice sets, and joint choices are determined by these.  %Imagine a simple setting where the choice functions of two decision makers are governed by the same underlying unobserved component, and, conditional on this component, the decision makers choose according to their choice function.   
We call such a joint choice function \emph{separable}.  At the most general level, we do not ask that the individual choice functions be rational in any sense. We imagine the stochastic analogue of this model:  choice data which captures the joint choices of the two decision makers. Thus, we study random joint choice rules, where for each pair of choice sets, we observe a distribution over joint choices. To the best of our knowledge, this type of data is novel to the theoretical stochastic choice model. However, this type of data is not novel to the field of econometrics and has seen much use there in applied work.\footnote{Most commonly, joint choice data has been used in the study of peer effects (see \citet{sacerdote2001peer}, \citet{kremer2008peer}, and \citet{card2013peer}). In addition, this type of data has been used to study the return on education by studying twins (see \citet{ashenfelter1994estimates} and \citet{miller1995twins}).} Our contribution here is to study random joint choice rules axiomatically.\footnote{ Another recent paper, \citet{dardanoni2020mixture}, studies a richer form of data. \citet{dardanoni2020mixture} studies probability distributions over choice functions. Our data considers probability distributions over alternative pairs from product sets.
}

If there is an unobserved component conditional on which joint choice is separable, then each individual must have a well-defined random choice rule defined via marginal distributions.  Formally, for each agent, we can define the probability they choose alternative $x$ from choice set $A$ independently of the choice set of the other decision maker.  We call this condition \textit{marginality}.  Marginality is a consequence of the separable model, but it also has normative appeal. In a situation where a decision maker's choices are only governed by the unobserved latent component, there is no reason that her marginal distribution of choices would depend on any other individual's behavior.  Thus, it is a minimal condition we would expect to see in the absence of peer effects or other types of direct influence.  To this end, our primary goal is to explore the relationship between separability and marginality.

%In order to study separability and marginality, we focus our analysis on the study of random joint choice rules. 

Our first result establishes a connection between marginality and separable joint choice rules.  %A separable choice rule is a deterministic joint choice function which can be decomposed into the product of single-agent choice functions.  Separable choice rules model separable choice without imposing rationality on either decision maker. 
We show that a random joint choice rule satisfies marginality if and only if it can be represented as the expectation of a signed measure over degenerate separable joint choice rules.  We also show that there are random joint choice rules satisfying marginality which cannot be represented as a probabilistic expectation of separable choice rules; thus in general, the signed measure obtained in the representation needs to take negative values. This tells us that the stochastic analogue of separability imposes further testable implications beyond marginality. Our second result shows that we may without loss restrict the support of our signed measure in the marginality characterization. A random joint choice rule satisfies marginality if and only if it can be represented as the expectation of a signed measure over rational separable choice functions:  essentially, pairs of linear orders. In other words, we can impose the requirement that choice behavior is rational in our signed measure and still recover any data set satisfying marginality.

Our third result focuses more on this multi-agent approach to random utility maximization:  we call this the separable random utility model.  We search for a probability measure over linear order pairs which induces the random joint choice rule. Maximization of each linear order dictates choice for each agent.  We establish that marginality plus an extension of the standard random utility axiom fail to characterize this model.\footnote{The standard random utility axiom is non-negativity of what are called the Block-Marschak polynomials. Non-negativity of the Block-Marschak polynomials should be thought of as the choice probabilities of each alternative satisfying a strong form of monotonicity on the set inclusion order. We extend this axiom to joint choice probabilities.} An extension of the counterexample used in our first result holds here. Our third result shows that if, in addition to marginality and the random utility axiom, we further impose that the marginal choices of either agent have a unique random utility representation, then the data can be represented by separable random utility.

An important practical implication of these results is that reliance on marginal data alone can be misleading.  The richer data introduced in our manuscript allows us a more powerful language to talk meaningfully about unobserved variables.  This data allows us a stronger test for falsifying the latent variable hypothesis.  To put this into a practical context:  imagine we only have marginal data available for each agent.  It would automatically appear as if each agent has their own stochastic choice function, responding to an  unobserved variable.  On the other hand, considering the joint choice structure can often lead us to falsify the hypothesis that the agents jointly respond to a latent variable.  This suggests a need for richer data involving the joint choice structure of multiple agents.  Similar remarks apply to the case of individuals who are apparently stochastically rational according to their marginals, but are not jointly stochastically rational.

%There are random joint choice rules whose marginal distributions exist, but where the joint data is not stochastically separable.  Thus, if joint choice data are available, even if marginal choice data exist for each agent, this does not imply that agents are actually behaving in response to a latent variable.  Similarly, there are random joint choice rules whose marginal distributions exist and appear stochastically rational, but where the joint data has no such rationalization.  Thus, testing for individual stochastic rationality with marginal choice data alone may lead to the incorrect implication that the individuals condition their preferences on some unobserved latent variable.  They may both pass the test as individuals, but if they do not pass this test jointly, there is no latent component that can explain both of their choices, and at least one or both of them must not conform to the model in question.

The rest of this paper is organized as follows. We conclude this section with a review of the related literature. Section 2 introduces preliminaries and discusses the type of data we consider. Section 3 discusses marginality and introduces our main results. Section 4 discusses separable random utility and our third result. Finally, we conclude in Section 5.

\subsection{Related Literature}

Our paper is related to the literature studying the random utility model of \citet{block1959random}. \citet{falmagne1978representation}, \citet{barbera1986falmagne}, and \citet{mcfadden1990stochastic} contribute to the literature by providing characterizations of random utility maximization. We add to this literature by considering a multiagent version of the random utility model. This strand of literature was revitalized with \citet{gul2006random} which considers an extension of random utility to randomization over expected utility functions. Notably, our paper is related to the subset of literature which studies dynamic random utility. The main formulation of \citet{frick2019dynamic} considers an extension of \citet{gul2006random} to a dynamic setting with preferences being separable over time. 

Most similar to our work is the work of \citet{li2021axiomatization}, which considers a similar environment to ours. Li considers a dynamic version of the standard random utility model with preferences being separable over time. The full characterization offered by Li is an extension of \citet{clark1996random}. Li also independently considers the same nonnegativity axiom as we have, and consequently we share some results with him.  Our focus is on marginality, whereas his primary contribution is to investigate the separable random utility model.  He proposes a partial characterization of this model using supermodularity of choice probabilities for a sufficiently small choice environment.

Further, our paper is related to the decision theory literature studying multi-agent choice. Much of this literature focuses on testing game theoretic models. \citet{sprumont2000testable} focuses on the testable implications of Nash equilibrium in normal form games, \citet{ray2001game} focuses on extensive form games, and \citet{lee2012testable} focuses on zero-sum games. \citet{carvajal2013revealed} takes a more focused approach and studies the testable implications of the Cournot model. More recently, the decision theoretic literature has begun to focus on testing models of social influence. \citet{cuhadaroglu2017choosing} and \citet{borah2018choice} consider deterministic models of social influence. \citet{ccm} and \citet{kashaev2019peer} consider stochastic models of social influence. The former studies an extension of the Luce model to multiple agents and the latter studies stochastic choice via random consideration.

Finally, our paper is also loosely related to the mathematical and information design literature on feasible joint posteriors. This literature studies when a posterior belief distribution is feasible for a group of agents who share a common prior. We study a decision theoretic analogue. We ask when a joint choice distribution can come from a single distribution over preference pairs. Similar to how we show that marginality is necessary but not sufficient for separability, this literature shows that the standard ``martingale condition'' (law of iterated expectations of beliefs) is necessary but not sufficient for feasibility. The mathematical literature began with \citet{dawid1995coherent} which characterizes the set of feasible distributions for two agents with two states of the world. \citet{arieli2021feasible} introduces this result to the economic literature and extends it to any number of agents. \citet{mathevet2020information} considers the question of feasibility through the lens of Bayesian persuasion and is able to offer an implicit characterization.

\section{The Model}\label{model}
\subsection{The Data}\label{data}
We take as given a pair of nonempty and finite sets of alternatives $X$ and $Y$. Let $\mathcal{X}$ and $\mathcal{Y}$ be the sets of all non-empty subsets of $X$ and $Y$, respectively. In this paper, we consider a data set which is novel in the context of decision theory. In this world, $(A,B)$ represents a decision problem where $A \in \mathcal{X}$ and $B \in \mathcal{Y}$. We assume that the outside observer has  a technology for observing the probability of $(x,y)$ being chosen from $(A,B)$, where $x \in A$ and $y \in B$. For each $(A,B)\in \mathcal{X} \times \mathcal{Y}$, $p(A,B)$ is a probability measure.\footnote{It should be understood that the notation $p(\cdot|A,B)$ does not refer to conditional probability.}

\begin{definition}
A \emph{random joint choice rule} is a map $p:\mathcal{X} \times \mathcal{Y}\rightarrow \bigcup_{(A,B)\in \mathcal{X} \times \mathcal{Y}}\Delta(A\times B)$ for which for each $(A,B)\in \mathcal{X} \times \mathcal{Y}$, we have $p(A, B)\in \Delta(A\times B).$\footnote{A random joint choice rule is notably weaker than the standard random choice rule for $X \times Y$. In the standard framework sets of the form $\{(a,x),(b,y)\}$ are observed. However, these types of sets are not observed in our data. We observe sets of the form $\{(a,x),(a,y),(b,x),(b,y)\}$.}
\end{definition}

We offer a few interpretations of this data set, though there are many more. We interpret $X$ as the global choice set from which agent $1$ may choose and $Y$ as the set from which agent $2$ may choose. Random joint choice data  represents the joint probability of agent $1$'s choices and  agent $2$'s choices from two sets, respectively. A reasonable example here might be the voting decisions of two senators. There are many other obvious examples. Education level choices among twins, choice of major among roommates, gardening choices of two neighbors, food choices of wife and husband, and sport activity choices of two siblings are some examples of random joint choice data of two distinct agents.

Another interpretation of our data is that they are the outcome of repeated choices by the same agent (intrapersonal). This single agent makes choices from $X \times Y$, which might depend on unobservable random factors. It is reasonable to ask if these choices on $X$ and $Y$ are independent, in the sense that consumption of a member of $X$ does not influence the preferences over $Y$.\footnote{Formally, one could test whether preferences over $X\times Y$ can be written with a utility representation of the form: $U(x,y)=H(f(x),g(y))$, where $H$ is increasing in each coordinate.} Under this interpretation, $p(x,y|A,B)$ represents the probability of the agent choosing $x$ and $y$ from $A$ and $B$, respectively, when offered a choice set of the form $A\times B$. By choosing from a choice set of form $A \times B$, we are implicitly assuming that there are no constraints across choice sets. In the abstract setting of our paper, this poses no issue, but issues may arise when we consider an agent who faces a set of prices for each good and is restricted by a total monetary budget. Alternatively, we may think of this single individual instead randomly choosing contingent choices or contracts (random choice of acts, see \emph{e.g.} \citet{lu2021random}).  $X$ and $Y$ now represent the available actions in two states of the world, state $1$ and state $2$, respectively.  The fact that the individual chooses from $X\times Y$ means that there are no interstate constraints.

A third interpretation of our data is that it captures population level choice data over time. In this interpretation, $X$ represents the set of alternatives available in the first time period and $Y$ represents the set of alternatives available in a second time period. Here, $p(x,y|A,B)$ represents the probability that an agent chooses $x$ in the first period and chooses $y$ in the second period. Just as was the case in the intrapersonal interpretation of our model, we are assuming here that there are no constraints on choice across time other than the observed choice sets. That is to say, an agent's choice in one period does not impact their choice set in another period.  \citet{frick2019dynamic}, \citet{duraj2018dynamic}, and \citet{li2021axiomatization} study dynamic stochastic choice using conditional random choice rules instead of random joint choice rules.\footnote{Dynamic stochastic choice has also been studied in the context of Luce's model or logit \citep{luce1959individual}, see in particular \citet{rust1987}, \citet{fudenberg2015dynamic}, \citet{lu2018random}, or \citet{pennesi2021intertemporal}.}

Our data allow us to study the correlation structure of stochastic choice across a pair of agents.\footnote{The theoretical framework can easily be generalized to accommodate any finite number of agents.}  Such a framework allows a richer language for discussing stochastic choice, and also allows for more restrictive testing. Table \ref{table:data} illustrates our data for two decision problems. For each decision problem, we also provide marginal choice distributions. Here, the marginal distribution of $\{x_1,x_2,x_3\}$ is the same across two choice problems: $(\{x_1,x_2,x_3\},\{y_1,y_2,y_3\})$ and $(\{x_1,x_2,x_3\},\{y_1,y_2\})$. This is not true in general. 

\begin{table}[h!]
    \centering
    \begin{tabular}{cc}
\toprule
{$ \begin{array}{c|ccc|c}
              & y_{1} & y_{2} & y_{3} &   \\
          \cline{1-5} 
x_1 & \multicolumn{1}{c}{0.2}           &  0           & 0.3             & 0.5 \\
x_2   & \multicolumn{1}{c}{0.1}           &  0.3           & 0            & 0.4     \\
x_3   & \multicolumn{1}{c}{0}           &  0           & 0.1            & 0.1       \\
\cline{1-5} 
 & 0.3           & 0.3  & 0.4 & 1   
\end{array}$} & {$\begin{array}{c|cc|c}
        & y_{1}  & y_{2}  &    \\
          \cline{1-4} 
x_1 & \multicolumn{1}{c}{0.3}           &  0.2                        & 0.5        \\
x_2   & \multicolumn{1}{c}{0.1}           &  0.3                      & 0.4         \\
x_3   & \multicolumn{1}{c}{0}           &  0.1                       & 0.1        \\
\cline{1-4} 
 & 0.4           & 0.6   & 1   
\end{array}$}\\
\bottomrule  \\
\end{tabular}

    \caption{Random joint choice rule for $(\{x_1,x_2,x_3\},\{y_1,y_2,y_3\})$ and $(\{x_1,x_2,x_3\},\{y_1,y_2\})$.}
    \label{table:data}
\end{table}

\subsection{Further Primitives}
Let $\mathcal{C}(X)$ denote the set of choice functions $c:\mathcal{X} \rightarrow X$ that satisfy $c(A) \in A$. Define $\mathcal{C}(Y)$ similarly.

\begin{definition}
A \textit{joint choice function} is a function $c:\mathcal{X} \times \mathcal{Y} \rightarrow X \times Y$ satisfying $c(A,B) \in A\times B$. 
\end{definition}

\begin{definition}
We say a joint choice function $c$ is \textit{separable} if there exists a choice function over $X$, $c_1$, and a choice function over $Y$, $c_2$, such that $c(A , B)= (c_1(A),c_2(B))$ for all $(A,B) \in \mathcal{X} \times \mathcal{Y}$.
\end{definition}

Separability reflects a basic normative criterion that we would expect joint choice functions to satisfy when there is a lack of direct interaction or complementarity between the two individual's choices.  Thus, it really reflects the idea that the two individuals act independently of each other.  Separability would often be ruled out, for example, in an environment where the pair of individuals constitute a household and there may be some complementarity between $X$ and $Y$.  It might also be ruled out when one agent is a leader and the other a follower or copycat:  For example, imagine a simple environment where $X = Y$ and agent $2$ is a leader, with choice function $c_2$.  Suppose also that agent $1$ has their own choice function $c_1$.  We can define a joint choice function $c(A, B)=(c_2(B), c_2(B))$ when $c_2(B)\in A$, and $c(A, B) = (c_1(A), c_2(B))$ otherwise.  Such a joint choice function typically violates separability. 

Let $\mathcal{C}(X) \times \mathcal{C}(Y)$ denote the set of separable choice rules over $X \times Y$. Let $\Delta(\mathcal{C}(X) \times \mathcal{C}(Y))$ denote the set of probability distributions over $\mathcal{C}(X) \times \mathcal{C}(Y)$ with typical element $\pi$. Let $\mathcal{L}(A)$ be the set of all linear orders over some set $A$.\footnote{A linear order is a binary relation which is complete, transitive, and antisymmetric.} For any set $A$, linear order $\succ \in \mathcal{L}(A)$, and menu $B \subseteq A$, define $M(B,\succ)$ as the unique maximal element of $B$ according to $\succ$. Let $\Delta(\mathcal{L}(X) \times \mathcal{L}(Y))$ denote the set of probability distributions over $\mathcal{L}(X) \times \mathcal{L}(Y)$ with typical element $\nu$.

\section{Stochastic Separability and Marginal Choice}

Consider the following property of a stochastic joint choice rule.

\begin{definition}
We say that a random joint choice rule is \textit{stochastically separable} if there exists some $\pi \in \Delta(\mathcal{C}(X) \times \mathcal{C}(Y))$ such that the following holds for all $(A,B)\in\mathcal{X}\times\mathcal{Y}$ and $(x,y) \in A\times B$.
    \begin{equation*}
        p(x,y|A,B) = \sum_{(c_1,c_2) \in \mathcal{C}(X) \times \mathcal{C}(Y)} \pi(c) \mathbf{1}\{c_1(A)=x,c_2(B)=y\}
    \end{equation*}
\end{definition}

Stochastic separability is the requirement that there is some latent variable governing joint choice behavior, but that conditional on this variable, choice is separable. 

Our main goal is to explore the relationship between random joint choice rules with well defined marginal random choice rules and random joint choice rules which are stochastically separable. In general, random joint choice rules do not have well defined marginal choice rules. We cannot define $p(x,A)$, the probability of choosing $x$ from $A$, independently of the choice set $B$.  The following property allows us to do so.

\begin{axiom}[Marginality]
For all $A\in \mathcal{X}$, $x\in A$, and all $B,B’\in \mathcal{Y}$, $\sum_{y\in B}p(x,y|A,B) = \sum_{y’\in B’}p(x,y'|A,B’)$, with a similar statement for $y\in B$.
\end{axiom}

A random joint choice rule has well defined marginal choice rules if and only if it satisfies marginality.

\begin{definition}
For a random joint choice rule $p$ satisfying marginality, define the \textit{marginal random choice rules} $p_1$ and $p_2$ as follows.
\begin{itemize}
    \item $p_1(x,A)=\sum_{y\in Y} p(x,y|A,Y)$
    \item $p_2(y,B)=\sum_{x\in X} p(x,y|X,B)$
\end{itemize}
\end{definition}

We are interested in marginality for several reasons. First, marginality is a necessary condition if one wants to test a model or axiom which imposes restrictions on marginal choices. Second, marginality has apparent normative appeal.  It is satisfied by every separable joint choice function, and the motivation behind it is much the same.  Indeed, it is clearly a necessary condition for stochastic joint choice to be stochastically separable.

%\begin{definition}
%We say that a random joint choice rule is \textit{stochastically separable} if there exists some $\pi \in \Delta(\mathcal{C}(X) \times \mathcal{C}(Y))$ such that the following holds for all $A \times B$ and $(x,y) \in A\times B$.
 %   \begin{equation*}
 %       p(x,y|A,B) = \sum_{c \in \mathcal{C}(X) \times \mathcal{C}(Y)} \pi(c) \mathbf{1}\{c(A,B)=(x,y)\}
  %  \end{equation*}
%\end{definition}

%In this finite setting, a data set is representable by a separable choice rule if and only if it is representable by maximization of a utility function of the form $U(x,y,A,B)=H(f(x,A),g(y,B))$ where $H$ is increasing in each agent. In this sense, separable choice rules capture separable choice without imposing classical rationality. With this in mind, stochastic separability exactly captures the random joint choice rules that are in the convex hull of separable choice rules. We now characterize the connection between marginality and separable choice rules.

The following result presents the first of two characterization theorems we provide involving stochastic separability.  While stochastic separability implies marginality, the converse fails.  However, marginality is in fact equivalent to a mathematical weakening of stochastic separability whereby the distribution of the latent variable is allowed to take negative values.  In other words, the class of random joint choice rules satisfying marginality coincides with the linear span of the separable joint choice rules.

In the statement of Theorem~\ref{marginalindependent}, a \emph{signed measure} over $\mathcal{C}(X)\times \mathcal{C}(X)$ is an element of $\pi\in\mathbb{R}^{\mathcal{C}(X)\times\mathcal{C}(Y)}$ for which $\sum_{c\in \mathcal{C}(X)\times\mathcal{C}(Y)}\pi(c)=1$.

\begin{theorem}\label{marginalindependent}
\begin{enumerate}
    \item A random joint choice rule $p$ satisfies marginality if and only if there exists a signed measure $\pi$ over $\mathcal{C}(X) \times \mathcal{C}(Y)$ such that for all $A \in \mathcal{X}$, $B \in \mathcal{Y}$, $x \in A$, and $y \in B$ we have the following.
    \begin{equation*}
        p(x,y|A,B) = \sum_{c \in \mathcal{C}(X) \times \mathcal{C}(Y)} \pi(c) \mathbf{1}\{c(A,B)=(x,y)\}
    \end{equation*}
    \item There exist random joint choice rules which satisfy marginality but are not stochastically separable.
\end{enumerate}
\end{theorem}

We leave all proofs to the appendix. This theorem tells us that marginality is a necessary but not sufficient condition for stochastic separability. The second part follows from the following example.

\begin{example}\label{ex:exampleseparable}Table~\ref{table:counterexample} below describes the main content of our counterexample. Here, $X = \{w,x,y,z\}$ and $Y=\{a,b,c,d\}$.  To save space, we only define the random joint rule on a few subsets; it is obvious that we can extend it to the remaining choice sets and preserve marginality.  We claim that this data is not compatible  with stochastic separability. Suppose that it is. Now consider $\{y,z\}\times \{a,b\}$. Only the pairs $(y,a)$ and $(z,b)$ are chosen here. Therefore, for any separable joint choice rule in the support, whenever $y$ is chosen from $\{y,z\}$, $a$ must be chosen from $\{a,b\}$.  Next, from the set $\{w,x\}\times \{a,b\}$, we can repeat this reasoning to establish that whenever $a$ is chosen from $\{a,b\}$ for a separable joint choice rule, it must be the case that $w$ is chosen from $\{w,x\}$.  Finally, inspecting $\{w,x\}\times\{c,d\}$ informs us that whenever $w$ is chosen from $\{w,x\}$, $c$ must be chosen from $\{c,d\}$.  Consequently, we infer that whenever $y$ is chosen from $\{y,z\}$, $c$ must be chosen from $\{c,d\}$.  However, the data from $\{y,z\}\times \{c,d\}$ actually requires the opposite:  whenever $y$ is chosen from $\{y,z\}$, $d$ must be chosen from $\{c,d\}$.  Therefore, Table~\ref{table:counterexample} cannot be compatible with stochastic separability. \end{example}

%To understand why the random joint choice rule is a counterexample, observe the following.  When choices are made from $\{a,b\} \times \{y,z\}$, only $(a,y)$ and $(b,z)$ are chosen. This means that for any separable choice rule in the support, whenever $a$ is chosen from $\{a,b\}$, it must be that $y$ is chosen from $\{y,z\}$. For choices from $\{a,b\} \times \{w,x\}$, we can apply the same logic to see that when $a$ is chosen from $\{a,b\}$, $w$ must be chosen from $\{w,x\}$. Following this logic one more time for $\{c,d\} \times \{w,z\}$ tells us that when where $c$ is chosen from $\{c,d\}$, it must be that $w$ is chosen from $\{w,z\}$. Putting all of these observations together, it should be the case that $c$ is chosen from $\{c,d\}$ if and only if $y$ is chosen from $\{y,z\}$. However, our random joint choice rule tells us the exact opposite happens at $\{c,d\} \times \{y,z\}$.

\begin{table}[h!]
    \centering
    \begin{tabular}{cc}
\toprule
{$ \begin{array}{c|cc}
 & a & b  \\
\cline{1-3} 
w & 0.5 & 0 \\
x & 0 & 0.5 \\

\end{array}$} & {$ \begin{array}{c|cc}
 & c & d  \\
\cline{1-3} 
w & 0.5 & 0 \\
x & 0 & 0.5 \\

\end{array}$}\\
\\
{$ \begin{array}{c|cc}
 & a & b  \\
\cline{1-3} 
y & 0.5 & 0 \\
z & 0 & 0.5 \\

\end{array}$} & {$ \begin{array}{c|cc}
 & c & d  \\
\cline{1-3} 
y & 0 & 0.5 \\
z & 0.5 & 0 \\

\end{array}$}\\
\bottomrule  \\
\end{tabular}

    \caption{Random joint choice rule which satisfies marginality but fails to be stochastically separable.}
    \label{table:counterexample}
\end{table}

The first part of Theorem \ref{marginalindependent} tells us that the random joint choice rules with well defined marginals are exactly those contained in the intersection of random joint choice rules and the linear span of separable choice rules. As the set of single agent choice functions is not linearly independent, the set of separable choice rules is also not linearly independent. A natural question is to what extent can we refine the set of separable choice rules while still characterizing marginality. We consider the refinement to rational separable choice rules.  

The signed measure $\nu$ in Theorem~\ref{marginalrational} refers to an element of $\mathbb{R}^{\mathcal{L}(X) \times \mathcal{L}(Y)}$ whose components sum to one.

\begin{theorem}\label{marginalrational}
A random joint choice rule $p$ satisfies marginality if and only if there exists a signed measure $\nu$ over $\mathcal{L}(X) \times \mathcal{L}(Y)$ such that for all $A \in \mathcal{X}$, $B \in \mathcal{Y}$, $x \in A$, and $y \in B$ we have the following.
\begin{equation*}
    p(x,y|A,B)=\sum_{(\succ, \succ')\in \mathcal{L}(X) \times \mathcal{L}(Y)} \nu(\succ,\succ')\mathbf{1}\{x=M(A,\succ),y=M(B,\succ')\}
\end{equation*}
\end{theorem}

Theorem~\ref{marginalrational} tells us that we can decompose any random joint choice rule with well defined marginals into a signed measure over rational separable choice rules. As a preliminary result necessary for the proof of this result, we show in the appendix, for a single-agent environment, the linear span of all rational choice functions coincides with all stochastic choice functions; thus, the linear span of the classical model has no testable content; see also \citet{dogan2022every}.  In contrast, our result establishes that signed measures over pairs of choice rules or linear order pairs are exactly characterized by marginality.

\section{Separable Luce}

Though marginality is not in general characteristic of stochastic separability, Theorem~\ref{marginalrational} indicates a close connection between the two.  Here, we study our framework in the context of the classical model of \citet{luce1959individual}.  We establish that the connection between marginality and stochastic separability is exact for the Luce model.

\begin{definition}
We say that a random joint choice rule $p$ is \textit{joint Luce} if there exists some $f:X\times Y \rightarrow\mathbb{R}_{++}$ such that the following holds for all $(A,B)\in \mathcal{X}\times\mathcal{Y}$ and all $(x,y) \in A\times B$.
    \begin{equation*}
        p(x,y|A,B) = \frac{f(x,y)}{\sum\limits_{a\in A,b\in B}f(a,b)}
    \end{equation*}
\end{definition}

Because $(X,Y)$ represents the pair of agents choosing from $X\times Y$, joint Luce rules can be characterized by the classical axioms of Luce.  When we think of $f(x,y)$ as a probability distribution over $X\times Y$, the joint Luce rule can be understood as a procedure whereby for any $(A,B)\in\mathcal{X}\times\mathcal{Y}$, the probability the element $(a,b)$ chosen is simply the conditional probability of that element, given the ``event'' $A\times B$ has obtained.  

Without positing a formal definition of influence, the general joint Luce model clearly allows for influence across the agents.  Consider a pair of agents conforming to the joint Luce model.  In this model, we can think of $f$ as a ``joint utility,'' or propensity to choose.  It is natural therefore to think that if no influence is present, we can decompose the joint propensity to choose into individual propensities to choose.  A natural way to do this in the joint choice context is to ask these propensities to choose be ``stochastically'' independent, when understood as probabilities.

%An outside observer witnessing the choice of the first agent typically learns something about the distribution of choices for the second agent.  That is, agent 1's choice is an informative signal of agent 2's choice, and conversely.  This suggests that at least one of the agents minimally takes into account the other agents' behavior when forming her own choice.

To this end, a natural model of separability and lack of influence in this framework is the following:

\begin{definition}
We say that a random joint choice rule is \textit{separable joint Luce} if there exist some $u:X\rightarrow\mathbb{R}_{++}$ and $v:Y\rightarrow\mathbb{R}_{++}$ such that the following holds for all $(A,B)\in\mathcal{X}\times\mathcal{Y}$ and $(x,y) \in A\times B$.
    \begin{equation*}
        p(x,y|A,B) = \frac{u(x)v(y)}{\sum\limits_{a\in A,b\in B}u(a)v(b)}
    \end{equation*}
\end{definition}

In the context of deliberate randomization, a separable joint Luce rule is one for which the agent's choices are stochastically independent.  Stochastic independence is a natural model of absolute ``lack of influence'' that we can write down without committing to a particular model of influence:  it posits that regardless of the choice of the first agent, we can say nothing whatsoever about the choice of the second agent.

Our main result concerning joint Luce rules is that amongst them, marginality is characteristic of stochastic separability.  Further, either of these two properties is equivalent to the separable Luce model.

\begin{proposition}\label{prop:luce}For joint Luce rules, the following are equivalent:
\begin{enumerate}
\item $p$ satisfies marginality 
\item $p$ is stochastically separable 
\item $p$ is a separable joint Luce rule.
\end{enumerate}
\end{proposition}

The takeaway here is that even though marginality is not characteristic of stochastic separability, in the context of the Luce rule, which is arguably the most commonly used model in single-agent stochastic choice, failures of marginality are actually characteristic of the presence of influence between the agents.

Importantly, Proposition~\ref{prop:luce} lends itself to methods of measuring ``behavioral influence.''  Identify a joint Luce rule with a probability measure on $X\times Y$:  that is, we can assume that for the joint Luce rule, the Luce weights $f$ are such that $\sum_{X\times Y}f(x,y)=1$.  Absence of influence (marginality) corresponds to a set $\mathcal{M}$ of probability measures on $X\times Y$ which are stochastically independent.  To this end, \emph{any metric} $d:\Delta(X\times Y)\times \Delta(X\times Y)$ can be used to define a natural measure of behavioral influence via $I(f)\equiv\inf_{f'\in\mathcal{M}}d(f,f')$.\footnote{In fact, an exercise like this can be done with any model violating rationality.  Here it is particularly natural as many metrics on the set of probability measures are already defined.}   We leave the study of such objects to future research.

\section{Separable Random Utility}

In the last section we considered stochastic separability in the context of the Luce model. In this section, we consider a more permissive model.

\begin{definition}
We say that a random joint choice rule is consistent with  \textit{separable random utility maximization} if there exists some $\nu \in \Delta(\mathcal{L}(X) \times \mathcal{L}(Y))$ such that the following holds for all $(A,B)\in\mathcal{X}\times\mathcal{Y}$ and $(x,y) \in A\times B$.
\begin{equation*}
    p(x,y|A,B)=\sum_{(\succ, \succ')\in \mathcal{L}(X) \times \mathcal{L}(Y)} \nu(\succ,\succ')\mathbf{1}\{x=M(A,\succ),y=M(B,\succ')\}
\end{equation*}
\end{definition}

In our setting, a data set is representable by a pair of linear orders if and only if it is representable by maximization of a utility function of the form $U(x,y)=H(f(x),g(y))$ where $H$ is increasing in each agent (a separable utility function). Separable random utility captures the convex hull of such utility functions. Our notion of separable random utility is an extension of the classic random utility model of \citet{block1959random} to a multiagent environment with separable preferences. The classic random utility model is characterized by non-negativity of the Block-Marschak polynomials. We introduce an analogue of the Block-Marschak polynomials for our environment.

\begin{definition}
For each $x \in A \subseteq X$ and $y \in B \subseteq Y$, the Block-Marschak polynomial, $q(x,y|A,B)$, is given by the following.\footnote{We note an important difference between the single agent Block-Marschak polynomials and the multi-agent Block-Marschak polynomials we consider here. In both cases, the Block-Marschak polynomials are the M\"{o}bius inverse (see \citet{rota1964foundations}) of choice probabilities. However, in the single-agent case the M\"{o}bius function is written as $(-1)^{|A' \setminus A|}$ while in the two-agent case the M\"{o}bius function is written as $(-1)^{|A' \setminus A| +|B'+B|}$ instead of $(-1)^{|(A' \times B') \setminus (A \times B)|}$. This follows from proposition 5 in \citet{rota1964foundations} and is a result of the product structure of our domain. Restating Rota's result, when considering a product set $A \times B$ and a partial order $R$ on $A \times B$, if $R$ is the product of a partial order $P$ on $A$ and a partial order $Q$ on $B$, then the M\"{o}bius function of $R$ on $A \times B$ is the product of the M\"{o}bius function of $P$ on $A$ and the M\"{o}bius function of $Q$ on $B$.}
\begin{equation*}\label{bmpolynomials}
\begin{split}
    q(x,y|A,B) & = \sum_{A':A \subseteq A'} \sum_{B':B \subseteq B'} (-1)^{|A' \setminus A| + |B' \setminus B|}p(x,y|A',B')\\
    & =p(x,y|A,B) - \sum_{A' \times B':A \times B \subsetneq A' \times B'}q(x,y|A',B')
\end{split}
\end{equation*}
\end{definition}

To interpret the Block-Marschak polynomials, consider the following. For some $(x,y)$, suppose each set $A\times B$ containing $(x,y)$ has a certain amount of probability it adds to the choice of $(x,y)$. The choice probability of $(x,y)$ from $A \times B$ is then given by the total probability added to $(x,y)$ be each weak superset of $A \times B$. The Block-Marshcak polynomials exactly capture how much probability each set $A \times B$ adds to the choice of $(x,y)$. This idea is captured by the recursive definition of the Block-Marschak polynomials above. Under this interpretation, non-negativity of the Block-Marschak polynomials corresponds to a strong form of monotonicity.

\begin{axiom}[Non-negativity]
For each $x \in A \subseteq X$ and $y \in B \subseteq Y$, $q(x,y|A,B) \geq 0$.
\end{axiom}

For random joint choice rules which satisfy marginality, when we decompose the data into a signed measure over linear order pairs, $q(x,y|A,B)$ is necessarily equal to the probability weight put on linear order pairs that rank $X \setminus A \succ x \succ A \setminus \{x\}$ and $Y \setminus B \succ y \succ B \setminus \{y\}$. Due to this relation, non-negativity of the Block-Marschak polynomials is a necessary condition for separable random utility maximization. It is natural to ask if non-negativity and marginality are sufficient conditions for separable random utility maximization. Example \ref{counterexamplerational} shows that this is not the case.

\begin{example}\label{counterexamplerational}
Let $X=\{a,b,c,d\}$ and $Y=\{w,x,y,z\}$. Consider the random joint choice rule induced by the following behavior. When the choice set is not a weak subset of $\{c,d\} \times \{y,z\}$, choices are made according to the following distribution over linear order pairs.
    \begin{equation*}
        \nu_1(\succ,\succ') =    \begin{cases}
            \frac{1}{2} \text{ }(a \succ b \succ c \succ d, w \succ' x \succ' y \succ' z)\\
            \frac{1}{2} \text{ }(b \succ a \succ d \succ c, x \succ' w \succ' z \succ' y)
        \end{cases}
    \end{equation*}
When the choice set is a weak subset of $\{c,d\} \times \{y,z\}$, choices are made according to the following distribution over linear order pairs.
    \begin{equation*}
        \nu_2(\succ,\succ') =    \begin{cases}
            \frac{1}{2} \text{ }(d \succ c, y \succ' z)\\
            \frac{1}{2} \text{ }(c \succ d,  z \succ' y) 
        \end{cases}
    \end{equation*}
The random joint choice rule described by the above behavior satisfies marginality and non-negativity.
\end{example}

To see why this is a counterexample, observe the following.  Suppose that the behavior is consistent with separable joint random utility and is rationalized by $\nu$.  By considering $\{a,b\}\times \{y,z\}$, the event $y \succ' z$ occurs if and only if the event $a \succ b$ occurs.  By considering $\{a,b\}\times \{w,x\}$, $a \succ b$ occurs if and only if $w \succ' x$ occurs.  And by considering $\{c,d\}\times\{w,x\}$, $w \succ' x$ occurs if and only if $c \succ d$ occurs.  So, $y\succ' z$ occurs if and only if $c \succ d$ occurs.  In other words, $\nu(\{\succeq,\succeq':c \succ d,z\succ' y\})=0$. However, by considering $\{c,d\}\times \{y,z\}$, the event $c \succ d$ occurs if and only if the event $z \succ' y$ occurs; and this must obtain with probability $.5$. This means that there is no probability distribution over linear order pairs that can induce the behavior described in Example \ref{counterexamplerational}. Also, note that the random joint choice rule described in Example~\ref{counterexamplerational} contains the behavior described in Example~\ref{ex:exampleseparable} and Table~\ref{table:counterexample}. Notably, this tells us that our counterexample fails to be separable despite satisfying marginality.

Our Example \ref{counterexamplerational} is related to the counterexample of \citet{fishburn1998stochastic} used to show that the random utility model is not identified. The marginal distributions over preferences induced by $\nu_1$ correspond to one of the distributions that Fishburn shows is observationally equivalent to some other distribution over preferences. This tells us that the marginal random choice rules from Example \ref{counterexamplerational} fail to have a unique random utility representation. We now show that the cases when marginality fails to guarantee separability are deeply connected to the cases when marginal random choice rules fail to have a unique random utility representation. 

Our first step in exploring this relation is to show that the marginal random choice rules we are considering actually have random utility representations. As we discussed earlier, marginality and non-negativity are necessary conditions for separable random utility maximization. These two axioms guarantee that the marginal random choice rules have a random utility representation.

\begin{lemma}
Suppose that a random joint choice rule $p$ satisfies non-negativity. Then the marginal random choice rules $p_1(x,A)$ and $p_2(y,B)$ each have a random utility representation.
\end{lemma}

The intuition behind this result is as follows. Given our specific definitions of $p_1$ and $p_2$, they are both well-defined even if marginality does not hold. It then follows that non-negativity of the random joint choice rule implies that the marginal random choice rules have non-negative Block-Marschak polynomials. This is necessary and sufficient for the existence of a random utility representation. Now that we have guaranteed that our marginal random choice rules have a random utility representation, we must now give conditions under which they have a unique random utility representation. \citet{turansick2022identification} gives necessary and sufficient conditions on the Block-Marschak polynomials which guarantee that a random choice rule has a unique random utility representation. We can check that our marginal random choice rules have a unique random utility representation using these conditions. We are now able to show the connection between failures of separability and failures of uniqueness in the marginals.

\begin{theorem}\label{uniqueconvex} 
Suppose a random joint choice rule $p$ satisfies marginality and at least one of its marginal random joint choice rules has a unique random utility representation. Then $p$ is consistent with separable random utility maximization if and only if it satisfies non-negativity.\footnote{To extend this result to more than two agents, it is required that all but one marginal random choice rule has a unique random utility representation. Further, Theorem 3 captures as a special case when $|X|\leq 3$ is satisfied for all but one agent.}
\end{theorem}

This theorem tells us that counterexamples like Example \ref{counterexamplerational} only exist when both marginal  random choice rules fail to have a unique random utility representation. The intuition behind this result is as follows. If one marginal random choice rule has a unique representation, then we can recover the random choice rule of the other agent conditional on each linear order of the first agent. Once we have these conditional random choice rules, marginality guarantees that the probabilities of these conditional random choice rules behave as if they are from single agent random choice rules. Non-negativity then guarantees that these conditional random choice rules have a random utility representation. Recall that this representation is conditional on a linear order for the first agent. We can then recover a distribution over linear order pairs by combining the marginal distribution over linear orders for the first agent with each conditional distribution over linear orders for the second agent.

\begin{remark}Separable random utility maximization is a special case of a more general model, whereby we envision a distribution over classically rational joint choice rules.  Say a joint choice rule is classically rational if there is a linear order $\succeq$ over $X\times Y$ such that for all $(A, B)\in\mathcal{X}\times\mathcal{Y}$, $c(A, B)=\arg\max_{\{(a,b)\in A\times B\}}\succeq$.  A finite set of linear inequalities characterization of this model in this restricted domain is unknown.  But, this model generally violates non-negativity:  for a simple example, let $X=\{a,b\}$, $Y=\{c,d\}$, where the preference is specified by:  $(b,d) \succ (a,c) \succ (a,d) \succ (b,c)$.  Then $q(a,c|\{a\}\times\{c\})= -1.$   We do not know if the conjunction of this model with marginality coincides with separable random utility maximization and leave this to future research.\footnote{We  thank an anonymous referee for suggesting we discuss the connection of this model with ours.}\end{remark}

\section{Conclusion}
In this paper we study random joint choice rules, a type of data that is commonly used to study peer effects. To our knowledge, this type of data is novel to the theoretical stochastic choice framework. Of primary interest is the connection between marginality and stochastic separability. We find that marginality is a necessary but  not sufficient condition for stochastic separability. Under the interpretation that our data arises from two decision makers, failures of stochastic separability might be thought of as one agent influencing the other. With this in mind, we can test for the presence of influence through failures of marginality. However, as marginality does not characterize stochastic separability, measuring the degree by which marginality fails does not necessarily serve as a sound measure of influence. Notably, the counterexample used in Theorem \ref{marginalindependent} satisfies marginality and yet attempting to decompose any portion of the data into separable choice rules leads to negative choice probabilities. In other words, if the data of this counterexample can be represented by a probability distribution over deterministic decision maker pairs, then at least one decision maker in each pair influences the other decision maker.

In addition to the results we provide, part of our contribution is in developing the tools to analyze random joint choice rules. Specifically, in the appendix below, we develop a graphical representation of random joint choice rules satisfying marginality. Our graphical construction can be thought of as an extension of \citet{fiorini2004short}. To our knowledge, \citet{fiorini2004short} was the first to use flow polytopes to study the random utility model and the linear order polytope. Recently, \citet{turansick2022identification} uses the graphical representation of \citet{fiorini2004short} to study uniqueness in the random utility model. \citet{doignon2022adjacencies} uses the equivalence between the linear order polytope and the flow polytope considered in \citet{fiorini2004short} to study the adjacency of linear orders in the linear order polytope. Many of the techniques used in these papers can be extended to our environment and graphical construction in order to study separable random utility.

While our focus is on the relation between marginality and separability, an obvious open question is how to characterize stochastic separability and separable random utility in our environment. We have shown that marginality is necessary for separability, non-negativity and marginality are necessary for separable random utility, and that testing for separable random utility becomes easier when either agent has a unique random utility representation. While both separability and separable random utility can be characterized by extensions of the axiom or revealed stochastic preference \citep{mcfadden1990stochastic} as well as through the use of cones rather than convex hulls \citep{kitamura2018nonparametric}, neither of these characterizations are finite or normative. Ultimately, random joint choice rules are understudied from an axiomatic perspective. Studying this type of data can offer insight into many practical applications, especially those considering history dependence across time or social influence across peers.

\appendix
\section{Graphical Construction}

In this section, we describe a graphical representation of random joint choice rules which satisfy marginality. The representation is called the marginal graph system and will be of use when proving Theorems \ref{marginalrational} and \ref{uniqueconvex}. Recall that when marginality is satisfied, we can defined the marginal random choice rules $p(x,A)$ and $p(y,B)$. With this we can define the single agent Block-Marschak polynomials.
\begin{equation*}
    \begin{split}
        q_1(x,A)  & = \sum_{A' : A \subseteq A'} (-1)^{|A' \setminus A|}p_1(x,A') \\
        & = p(x,A) - \sum_{A \subsetneq A'} q_1(x,A')
    \end{split}
\end{equation*}
We define the Block-Marschak polynomials for the second agent, $q_2(y,B)$, similarly.

We now begin with the construction of the marginal graph system. For two agents, there are two different marginal graph systems; one for each ordering of the agents. We will focus on the construction where we focus on the marginal choices of the first agent and the conditional choices of the second agent. The marginal graph system consists of a collection of graphs which fit into two categories. The first category consists of a single graph which captures information about the marginal choices of the first agent. The second category consists of many graphs, each of which captures information about the choices of the second agent conditional on the choices of the first agent.

The construction of the graph corresponding to the first agent will follow closely with the construction of the graph considered in \citet{fiorini2004short}. We will call this graph the \textit{marginal component} of the marginal graph system. To begin, the nodes of the marginal component are indexed by the elements of $2^X$, the power set of $X$. We will refer to nodes by their indexed set. There exists an edge between nodes $A$ and $B$ if one of the following two conditions hold.
\begin{enumerate}
    \item $A \subseteq B$ and $|B \setminus A| = 1$
    \item $B \subseteq A$ and $|A \setminus B| = 1$
\end{enumerate}
For an edge connecting sets $A$ and $A \setminus \{x\}$, we assign $q_1(x,A)$ as the edge capacity.

We now consider the construction of the graphs corresponding with choices in the second agent. We will call this collection of graphs the \textit{conditional component} of the marginal graph system. Further, we call a single graph of the conditional component a \textit{conditional graph}. For each pair $(x,A)$ with $x \in A \subseteq X$, we construct a graph for the second agent. The constructions are analogous. Here we consider the construction for $(x,A)$. To begin, the nodes of a conditional graph are indexed by the elements of $2^Y$, the power set of $Y$. We will refer to nodes by their indexed set. There exists an edge between nodes $B$ and $C$ if one of the following two conditions hold.
\begin{enumerate}
    \item $C \subseteq B$ and $|B \setminus C| = 1$
    \item $B \subseteq C$ and $|C \setminus B| = 1$
\end{enumerate}
For an edge connecting sets $B$ and $B \setminus \{y\}$, we assign $q(x,y|A,B)$ as the edge capacity. We are capturing conditional choices here as we are using $q(x,\cdot|A,\cdot)$ to assign edge capacities.

We are interested in the representation of linear order pairs in terms of the marginal graph system. A linear order pair will be represented by a single path along the marginal component and a single common path along each conditional graph of the conditional component. We now consider the representation for $(\succ_1,\succ_2)$. The path on the marginal component corresponding to the preference $\succ_1$ has nodes satisfying the following.
\begin{itemize}
    \item $A$ such that for all $x \in X \setminus A$, $x \succ_1 M(A,\succ_1)$
\end{itemize}
The edges connecting these nodes will have capacities of form $q_1(M(A,\succ_1),A)$. The path on the conditional graphs corresponding to the preference $\succ_2$ has nodes satisfying the following.
\begin{itemize}
    \item $B$ such that for all $y \in Y \setminus B$, $y \succ_2 M(B,\succ_2)$
\end{itemize}
For a given $(x,A)$, the edges connecting these nodes will have capacities of form $q(x,M(B,\succ_2)|A,B)$.

\section{Preliminary Results}
\subsection{Lemma 1}
\begin{proof}
We show that each marginal Block-Marschak polynomials, $q_1(x,A)$ and $q_2(y,B)$, are non-negative. By \citet{falmagne1978representation}, this is sufficient for the existence of a random utility representation.
\begin{equation*}
    \begin{split}
        q_1(x,A) & = p(x,A) - \sum_{A \subsetneq A'} q_1(x,A') \\
        & = \sum_{A \subseteq A'} (-1)^{|A' \setminus A|} p(x,A) \\
        & = \sum_{A \subseteq A'} (-1)^{|A' \setminus A|} \sum_{y \in Y} p(x,y|A,Y) \\
        & = \sum_{y \in Y} \sum_{A \subseteq A'} (-1)^{|A' \setminus A|}  p(x,y|A,Y) \\
        & = \sum_{y \in Y} q(x,y|A,Y) \geq 0
    \end{split}
\end{equation*}
\end{proof}

\subsection{Lemma 2}
We begin by proving a preliminary lemma that will be used in the proof of Theorem \ref{marginalrational} and Theorem \ref{uniqueconvex}. This lemma consists of showing that marginality is equivalent to this next definition.

\begin{definition}
We say that a random joint choice rule satisfies \textit{recursivity} if for every $A \neq X$, every $B\subseteq Y$ and every $y\in B$, the following is satisfied (with a similar statement for sums across $B \subsetneq Y$).
$$\sum_{x\in A}q(x,y|A,B) = \sum_{z\in X\setminus A}q(z,y|A\cup \{z\},B)$$ 
\end{definition}

\begin{lemma}\label{lem:margrec}
The random joint choice rule $p$ satisfies marginality if and only if the corresponding Block-Marschak polynomials satisfy recursivity.
\end{lemma}

\begin{proof}
First we show that marginality implies recursivity. Fix $A,B$ and $y\in B$.  Let us write the equations:

\begin{equation}\label{eq:first}
\sum_{x\in A}q(x,y|A,B) = \sum_{x\in A}\sum_{A\subseteq A'}\sum_{B\subseteq B'}(-1)^{|A'\setminus A|+|B'\setminus B|}p(x,y|A',B' ).
\end{equation}

Likewise,
\begin{equation}\label{eq:second}
\sum_{z\in X\setminus A}q(z,y|A\cup \{z\},B)=\sum_{z\in X\setminus A}\sum_{A\cup \{z\}\subseteq A'}\sum_{B\subseteq B'}(-1)^{|A'\setminus (A \cup \{z\})|+|B'\setminus B|}p(z,y|A' \cup \{z\},B' ).
\end{equation}

Let us subtract equation~\eqref{eq:second} from equation~\eqref{eq:first}.  We will do a simple counting argument.  In particular, for every set $A' $, we will count the number of times it appears in the difference of the two equations.  In equation~\eqref{eq:second}, no term of the type $p(x,y|A,B' )$ ever appears.  Consequently, the difference of equation~\eqref{eq:first} and equation~\eqref{eq:second} has a term of $\sum_{x\in A}\sum_{B\subseteq B' }(-1)^{|B' \setminus B|}p(x,y|A,B' )$.

Now consider any set $A' $ for which $A \subseteq A' $ and $|A|<|A' |$.  The total coefficient coming from equation~\eqref{eq:first} is obviously
$$\sum_{x\in A}\sum_{B\subseteq B' }(-1)^{|A' \setminus A|+|B' \setminus B|}p(x,y|A' ,B' ).$$
Likewise, the total coefficient coming from the negation of equation~\eqref{eq:second} is $(-1)\sum_{z\in A' \setminus A}\sum_{B\subseteq B' }(-1)^{|A' \setminus A|-1+|B' \setminus B|}p(z,y|A' ,B' )$, or 
$$\sum_{z\in A' \setminus A}\sum_{B\subseteq B' }(-1)^{|A' \setminus A|+|B' \setminus B|}p(z,y|A' ,B' ).$$
Overall, then, the difference of equation~\eqref{eq:first} and equation~\eqref{eq:second} is 
$$\sum_{A\subseteq A' }\sum_{B\subseteq B' }(-1)^{|A' \setminus A|+|B' \setminus B|}\sum_{x\in A' }p(x,y|A' ,B' ).$$
Reverse the order of the sums according to $A$ and $B$, and obtain:
$$\sum_{B\subseteq B' }\sum_{A\subseteq A' }(-1)^{|A' \setminus A|+|B' \setminus B|}\sum_{x\in A' }p(x,y|A' ,B' ).$$
Now, by marginality, $\sum_{x\in A' }p(x,y|A' ,B' )$ is independent of $A' $.  Since $y\in B$ is fixed, we can call this term $\iota(B' )$.  Therefore the expression becomes:
$$\sum_{B\subseteq B' }\sum_{A\subseteq A' }(-1)^{|A' \setminus A|+|B' \setminus B|}\iota(B' ).$$
One more rearrangement:
$$\sum_{B\subseteq B' }(-1)^{|B' \setminus B|}\iota(B' )\sum_{A\subseteq A' }(-1)^{|A' \setminus A|}.$$
Obviously, though, since $A\neq X$, we know that $\sum_{A\subseteq A' }(-1)^{|A' \setminus A|}=0.$  This follows, as in the notation of Rota (1964), as what we have is $\sum_{A\subseteq A' }\mu(A,A' )\zeta(A' ,X)$ (here $\mu$ is the M\"{o}bius function of set inclusion), so that the entire expression is $\delta(A,X)$, where again this is the ``Kronecker delta" referred to in Rota (1964) as being the identity element of the incidence algebra.  This identity element is $0$ if $A \neq X$ (otherwise is 1).  So we are done.

Now we show that recursivity implies marginality. We begin with a claim. 
\begin{claim}\label{claim:recursion}
Suppose $q$ satisfies recursivity.  Then for each $n\geq 0$, we get $$\sum_{x\in A}\sum_{A’:A\subseteq A’:|A’|\leq|A|+n}q(x,y|A’,B) = \sum_{A’:A\subseteq A’:|A’|=|A|+n}\sum_{x \in A'}q(x,y|A’,B).$$ Further, $$\sum_{x\in A}\sum_{A’:A\subseteq A’}q(x,y|A’,B) = \sum_{z\in X}q(z,y|X,B).$$
\end{claim}

\begin{proof}[Proof of Claim~\ref{claim:recursion}]The proof is by induction on $n$.  For $n=0$ it is trivial. Suppose it is true for $n$ and we will show for $n+1$. This gives us 
\begin{equation*}
    \begin{split}
    \sum_{x\in A}\sum_{A’:A\subseteq A’:|A’|\leq |A|+n+1}q(x,y|A’,B) & = \sum_{x\in A}\sum_{A’:A\subseteq A’:|A’|\leq |A|+n}q(x,y|A’,B) \\
    & + \sum_{x\in A}\sum_{A’:A\subseteq A’:|A’|=|A|+n+1}q(x,y|A’,B).
\end{split}
\end{equation*}
By the induction hypothesis, this gives 
\begin{equation*}
    \begin{split}
        \sum_{x\in A}\sum_{A’:A\subseteq A’:|A’|\leq |A|+n+1}q(x,y|A’,B) & = \sum_{A’:A\subseteq A’:|A’|=|A|+n}\sum_{x\in A’}q(x,y|A’,B) \\
        & + \sum_{x\in A}\sum_{A’:A\subseteq A’:|A’|=|A|+n+1}q(x,y|A’,B)
    \end{split}
\end{equation*}
By applying recursivity to the first part of this sum, we know that $$\sum_{x\in A’}q(x,y|A’,B) = \sum_{z\in X\setminus A’}q(z,y|A’\cup\{z\},B).$$  So $$\sum_{A’:A\subseteq A’:|A’|=|A|+n}\sum_{x\in A’}q(x,y|A’,B) = \sum_{A’:A\subseteq A’:|A’|=|A|+n}\sum_{z\in X\setminus A’}q(z,y|A’\cup\{z\},B).$$  Now let us do a simple combinatorics argument.

Let us now consider a set $A^*$ of cardinality $|A|+n+1$, which contains $A$.  This set appears in the form of $A’\cup \{z\}$ in the above summation exactly $n+1$ times, one for each members $z\in A^*\setminus A$.  And each time, it adds a value of $q(z,y|A^*,B)$.  So, overall,

$$ \sum_{A’:A\subseteq A’:|A’|=|A|+n}\sum_{z\in X\setminus A’}q(z,y|A’\cup\{z\},B)=\sum_{A’:A\subseteq A’:|A’|=|A|+n+1}\sum_{x\in A’\setminus A}q(x,y|A’,B).$$  So the expression: $$\sum_{A’:A\subseteq A’:|A’|=|A|+n}\sum_{x\in A’}q(x,y|A’,B) + \sum_{x\in A}\sum_{A’:A\subseteq A’:|A’|=|A|+n+1}q(x,y|A’,B)$$
gives exactly $$\sum_{A’:A\subseteq A’:|A’|=|A|+n+1}q(x,y|A’,B),$$ which is what we wanted to prove in regards to the first part of the lemma. The second part of the lemma follows from setting $n= |X| - |A|$.
\end{proof}
Now observe that $\sum_{x\in A}p(x,y|A,B)=\sum_{B\subseteq B’}\sum_{x\in A}\sum_{A\subseteq A’}q(x,y|A,B)$, which by the above claim is the same as $\sum_{B\subseteq B’}\sum_{x\in X}q(x,y|X,B)$, which is independent of $A$. So we are done.
\end{proof}

\subsection{Lemma 3}
We now state and prove another preliminary lemma which will be used in the proofs of Theorem \ref{marginalrational} and Theorem \ref{uniqueconvex}. The statement of the lemma requires some preliminary notation. Fix $A \times B \subseteq X \times Y$ with $(x,y)\in A \times B$. Let
\begin{multline*}
    N_{(x,y),A \times B} = \\ \{(\succ_1,\succ_2):\forall (a,a',b,b')\in A^c\times A\times B^c \times B, a\succ_1 x \succ_1 a'\mbox{ and }b\succ_2 y \succ_2 b'\}.
\end{multline*}
Then $N_{(x,y),A\times B}$ consists of the set of order pairs $(\succ_1,\succ_2)$ for which:
\begin{enumerate}
\item Every member of $A$ is $\succ_1$ ranked below every member outside of $A$, where $x$ is at the top of $A$ 
\item Every member of $B$ is $\succ_2$ ranked below every member outside of $B$, where $y$ is at the top of $B$.
\end{enumerate}
The following lemma and its proof are analogous to the classical single agent case, see \emph{e.g.} \citet{falmagne1978representation}.
\begin{lemma}\label{lem:check}
The random joint choice rule $p$ is rationalized by a signed measure $\nu$ over $\call(X) \times \call(Y)$ if and only if $q(x,y|A,B) = \nu(N_{(x,y),A \times B})$ for all $(x,y,A,B)$.
\end{lemma}
\begin{proof}
For all $A\subseteq X$, all $B\subseteq Y$, all $x \in A$, and all $y \in B$, $\nu$ rationalizes the random joint choice rule if and only if $p(x,y|A,B)$ is the $\nu$-probability of realizing a pair $(\succ_1,\succ_2)$ for which $A \subseteq \{z\in X:x \succ_1 z\}$ and $B \subseteq \{w\in Y:y \succ_2 w\}$.
Namely, it is the $\nu$-probability of $\bigcup_{A\subseteq A'}\bigcup_{B\subseteq B'}N_{(x,y),A'\times B'}$.  For a fixed $(x,y)$, the sets $N_{(x,y),A'\times B'}$ are disjoint as $A\subseteq A'$ and $B\subseteq B'$, conclude:
\[p(x,y|A,B)=\sum_{A':A\subseteq A'}\sum_{B':B\subseteq B'}\nu(N_{(x,y),A'\times B'}).\] 
The result now follows from the M\"{o}bius inversion formula.
\end{proof}

Let $N(x,A)=\{\succ|X \setminus A \succ x \succ A \setminus \{x\}\}$. The following lemma holds by an analogous argument.

\begin{lemma}
A random choice rule is rationalizable by a signed measure $\nu$ over linear orders if and only if $q(x,A) = \nu(N(x,A))$ for all such $x \in A \subseteq X$.
\end{lemma}

\subsection{Lemma 5}

\begin{lemma}
Every random choice rule over $X$ can be expressed as a signed measure over linear orders of $X$.
\end{lemma}

This lemma was also proven as Theorem 2 in \citet{dogan2022every}. We provide an alternate proof here both for completeness and because we use the specific construction in our proof in later proofs.

\begin{proof}
Consider the standard probability flow diagram introduced by Fiorini (2004). We know from Theorem 3 of Falmagne (1978) that inflow equals outflow on this graph regardless of the (classically stochastic) rationalizability of the random choice rule. However, in our case some of these flows may be negative flows as there may be negative Block-Marschak polynomials. We will now use two algorithms to define two functions from $\mathcal{L}(X)$ to $\mathbb{R}$.

\begin{enumerate}
    \item Initialize at $i = 0$. Set $q_i(x,A)=q(x,A)$ and $g(\succ)=0$.
    \item If each $q_i(x,A)\geq 0$, set $l(x,A)=q_i(x,A)$ and terminate the algorithm. If not proceed to step 3.
    \item There exists some $q_i(x,A) < 0$. Choose a minimal one (i.e. most negative one). Call this minimal value $r$. This $q_i(x,A)$ is associated with some path from $X$ to $\emptyset$. Fix one of these paths. This path is bijectively associated with linear order $\succ_i$.
    \item For each edge constraint $q_i(x,A)$ along the fixed path, set $q_i(x,A)=q_i(x,A) - r$ (since r is negative). Set $g(\succ_i)=r$. Set $i = i+1$. Return to step 2.
\end{enumerate}

The above algorithm terminates when each $q_i$ is nonnegative. This means that each $l(x,A)$ is nonnegative. Since we know that subtracting out a path maintains inflow equals outflow, a probability flow diagram populated by these $l(x,A)$ also satisfies inflow equals outflow. For the next algorithm, consider a probability flow diagram populated by these $l(x,A)$.

\begin{enumerate}
    \item Initialize at $i=0$. Set $l_i(x,A) = l(x,A)$.
    \item If each $l_i(x,A) = 0$, terminate the algorithm. If not proceed to step 3.
    \item Since inflow equaled outflow at step 1 and inflow equals outflow is maintained throughout this algorithm, $l_i(x,A) > 0$ implies that there is some path from $X$ to $\emptyset$ with strictly positive edge capacities. Fix this path. Let $r$ be the minimum edge capacity along this path. Let $\succ_i$ be the linear order bijectively associated with this path.
    \item Set $h(\succ_i)=r$. For each edge capacity on this fixed path, set $l_i(x,A)= l_i(x,A)-r$. Set $i=i+1$. Return to step 2.
\end{enumerate}

This algorithm terminates with each $l_i(x,A)=0$. Now define the function $f:\mathcal{L}(X) \rightarrow \mathbb{R}$ as $f(\succ) = g(\succ) + h(\succ)$. Recall the set $N(x,A)= \{\succ|X \setminus A \succ x \succ A \setminus \{x\} \}$. By the construction of the the two algorithms (since the second algorithm terminates with zero flow and inflow equals outflow is maintained during the two algorithms), we know for the constructed $f$ that $f(N(x,A))=q(x,A)$. Thus $f$ rationalizes the arbitrary random choice rule $p$ by Lemma 4.
\end{proof}

Later on, we will use the terminology \textbf{decompose} a conditional graph. By this we mean that we run the two algorithms used in the prior proof on the conditional graph and return the function $f$.

\section{Omitted Proofs}

\subsection{Theorem \ref{marginalindependent}}
For the first part of the theorem, we begin with necessity. Separable choice rules individually satisfy marginality. It is then trivial to show that a linear combination of separable choice rules satisfy marginality. For sufficiency, observe that maximization of linear order pairs induce separable choice rules. Thus the sufficiency in Theorem \ref{marginalrational} implies sufficiency here. We leave the proof of sufficiency in Theorem \ref{marginalrational} to the next section.

For the second part of the theorem, the counterexample from Example \ref{counterexamplerational} acts as a counterexample here. The choice probabilities can be recovered from the description in the example. For completeness, Table \ref{tab:table1} provides the Block-Marschak polynomials of the counterexample.

\begingroup
\begin{table}[htbp]
\centering
\makebox[\linewidth]{
\begin{tabular}{c|c|c|c|c|c|c}
\hline\hline
 Set  & $\{w,x,y,z\}$ & $\{x,y,z\}$ & $\{w,y,z\}$ & $\{y,z\}$ & $\{y\}$ & $\{z\}$\\
 \hline   $\{a,b,c,d\}$ & $q(a,w)$ & $q(a,x)$ & $q(b,w)$ & $q(a,y)$ & $q(b,y)$ & $q(a,z)$ \\
 & $q(b,x)$ & & & $q(b,z)$ & & \\
 \hline $\{b,c,d\}$ & $q(b,w)$ & $q(b,x)$ & - & $q(b,y)$ & - & $q(b,z)$ \\
 \hline $\{a,c,d\}$ & $q(a,x)$ & - & $q(a,w)$ & $q(a,z)$ & $q(a,y)$ & - \\
 \hline $\{c,d\}$ & $q(c,w)$ & $q(c,x)$ & $q(d,w)$ & $q(c,z)$ & $q(c,y)$ & $q(d,z)$ \\
 & $q(d,x)$ & & & $q(d,y)$ & & \\
 \hline $\{c\}$ & $q(c,x)$ & - & $q(c,w)$ & $q(c,y)$ & - & $q(c,z)$ \\
 \hline $\{d\}$ & $q(d,w)$ & $q(d,x)$ & - & $q(d,z)$ & $q(d,y)$ & - \\
 \hline\hline 
\end{tabular}
}
\vspace{.2cm}
\caption {\label{tab:table1} 
Block-Marschak polynomial values for the counterexample. The corresponding sets for the first agent are given by the set column and the corresponding sets for the second agents are given by the set row. Each cell indicates the non-zero BM-polynomials of the random joint choice rule. All the non-zero BM-polynomials are equal to $0.5$  }
\end{table}
\endgroup

\subsection{Theorem \ref{marginalrational}}

\begin{proof}
Necessity is trivial as the linear combination of vectors satisfying marginality (an equality constraint) also satisfies marginality. Consider the marginal graph system representation described in Appendix A with $X$ being represented on the marginal graph and $Y$ being represented on the conditional graphs. Consider the following algorithm which takes in a marginal graph representation and returns a rationalizing signed measure.
  
\begin{enumerate}
    \item Initialize at $j=3$. Let $f(\succ,\succ')=0$.
    \item Choose some $(x,A)$ with $A\subseteq X$, $|A|=2$ and $x \in A$ such that there exists $y\in B \subseteq Y$ with $q(x,y|A,B) \neq 0$. Call $A = \{x,z\}$. If none exist, proceed to step (4).
    \item Decompose using the construction in the proof of Lemma 5 the conditional graph of $q(x,A)$. This decomposition returns a function $g:\mathcal{L}(Y) \rightarrow \mathbb{R}$. Fix $\succ \in \mathcal{L}(X)$ which ranks $z$ last and $x$ second to last. Set $f(\succ,\succ')=f(\succ,\succ')+g(\succ')$. Further, for each $\succ' \in \mathcal{L}(Y)$, set $q(w,y|C,B)=q(w,y|C,B)-g(\succ')$ if $X \setminus C \succ w \succ C \setminus \{w\}$ and $Y \setminus B \succ' y \succ' B \setminus \{y\}$ and maintain $q(w,y|C,B)$ otherwise. Return to step (2).
    \item For each $A \subseteq X$ with $|A|=2$, $x\in A$, $B \subseteq Y$, and $y \in B$ we have $q(x,y|A,B)=0$. Marginality is maintained at each step of this algorithm (as we are subtracting/adding terms which themselves satisfy marginality). This means the following holds for each $x\in X$, $B \subseteq Y$, and $y\in B$.
    \begin{equation*}
        \sum_{z \neq x} q(x,y|\{x,z\},B)=q(x,y|\{x\},B)
    \end{equation*}
    Since each term on the left side equals zero, it is the case that the right hand term is zero.
    \item Consider pairs of the form $(x,A)$ with $x \in A$ and $|A|=j$. Partition the collection of such pairs such that $(x,A)$ and $(z,C)$ are in the same partition if and only if $A \setminus \{x\}=C \setminus \{z\}= D$. Let $\mathcal{D} = \{(x,A)|x \in A, A \setminus \{x\} = D\}$. Consider some maximal collection of linear orders which rank the elements of $D$ lowest and agree on their ranking of $D$. Call this set $\mathcal{L}_D$.
    \item Choose some $(x,A)$ with $A\subseteq X$, $|A|=j$ and $x \in A$ such that there exists $y\in B \subseteq Y$ with $q(x,y|A,B) \neq 0$. If none exist, and $j<|X|-1$, set $j=j+1$ and return to step 5. If none exist and $j=|X|-1$, terminate the algorithm.
    \item Decompose the conditional graph of $q(x,A)$. This decomposition returns a function $g:\mathcal{L}(Y) \rightarrow \mathbb{R}$. Fix $\succ \in \mathcal{L}_D$ such that $X \setminus A \succ x \succ D$ and where $D = A \setminus\{x\}$. Set $f(\succ,\succ')=f(\succ,\succ')+g(\succ')$. Further, for each $\succ' \in \mathcal{L}(Y)$, set $q(w,y|C,B)=q(w,y|C,B)-g(\succ')$ if $X \setminus C \succ w \succ C \setminus \{w\}$ and $Y \setminus B \succ' y \succ' B \setminus \{y\}$ and maintain $q(w,y|C,B)$ otherwise. Return to step (6).
\end{enumerate}

We now argue that the above algorithm terminates with $q(x,y|A,B)=0$ for all such $(x,y,A,B)$. Suppose we are at the point in the algorithm where pairs of the form $(x,A)$ with $|A|=j$ are being considered. Suppose we know that for each $D$ with $|D|<j$, $q(x,y|D,B)=0$. Recall that, by Lemma 2, marginality is equivalent to recursivity. By recursivity, this tells us the following.
$$0=\sum_{z\in D}q(z,y|D,B) = \sum_{x\in X\setminus D}q(x,y|D\cup \{x\},B)$$ 
Now note that pairs of the form $(x,D \cup \{x\})$ are exactly the pairs that define the partition described in step (5) for the set $D$. This means that when we decompose (see the proof of Lemma 5) the conditional graph of each $(x,A)$ in $\mathcal{D}$, we add net zero weight to the collection of linear order pairs $(\succ,\succ')$ which satisfy $\succ \in \mathcal{L}_D$ and $Y \setminus B \succ' y \succ' B \setminus \{y\}$. Since each $\succ \in \mathcal{L}_D$ agree on their ranking of $D$, there exists a single $z \in D$ such that $q(z,y|D,B)$ is altered in this stage of the algorithm. This means that for each $w \neq z$, $q(w,y|D,B)=0$ as it was equal to zero before this stage of the algorithm. Further, since a net zero weight was added to the collection of linear order pairs $(\succ,\succ')$ which satisfy $\succ \in \mathcal{L}_D$ and $Y \setminus B \succ' y \succ' B \setminus \{y\}$, $q(z,y|D,B)$ faces a net zero change of weight during this stage of the algorithm. These two points together tell us that when we decompose the conditional graphs of $(x,A)$ with $|A|=j$, we end with $q(z,y|C,B)=0$ for all $C$ with $|C| \in \{j,j-1\}$. Recall that each linear order in $\mathcal{L}_D$ agrees on their ordering of $D$. By the same logic we applied for sets of size $j-1$, this means that each $q(z,y|C,B)$, with $|C|<j$, faces a net zero change during the stage of the algorithm when we consider sets of size $j$. As we started the algorithm by setting $q(x,y|A,B)=0$ for all $A$ with $|A| \leq 2$, this means that when the algorithm terminates, $q(x,y|A,B)=0$ for all $A$ with $|A|<|X|$. Finally, by marginality being equivalent to recursivity (Lemma 2), we know the following.

\begin{equation*}
    0=\sum_{z \neq x} q(z,y|X \setminus \{z\},B)= q(x,y|X,B)=0
\end{equation*}

Thus, the algorithm terminates with $q(x,y|A,B)=0$ for all such $(x,y,A,B)$. This finally means that the $f$ function produced by the algorithm satisfies the following.
\begin{equation*}
    q(x,y|A,B)=\sum_{(\succ, \succ')\in \mathcal{L}(X) \times \mathcal{L}(Y)} f(\succ,\succ')\mathbf{1}\{\succ=N(x,A),\succ'=N(y,B)\}
\end{equation*}
By Lemma 3, the signed measure constructed by our algorithm rationalizes the random joint choice rule. This is what we set out to prove and so we are done.

\end{proof}

\subsection{Proposition \ref{prop:luce}}

\begin{proof}Let us show that a separable joint Luce rule is stochastically separable.  Let $u,v$ be the functions associated with the joint choice rule; without loss, let us assume that $1=\sum_x u(x)=\sum_y v(y)$.  Consider the single-agent Luce rule on $X$ defined by $u$, and the single-agent Luce rule on $Y$ defined by $v$.  Call these $p_u$ and $p_v$ respectively.  It is known that $p_u$ is induced by a probability distribution $\pi_u$ on $\mathcal{C}(X)$ and $p_v$ by a distribution $\pi_v$ on $\mathcal{C}(Y)$.\footnote{This true for any single-agent stochastic choice function, in particular the distributions $\pi_u$ and $\pi_v$ in this case can be taken to be distributions over classically rational choice functions, see \emph{e.g.} \citet{BMidentity,debreuidentity}.}  Now let $\pi$ be the distribution over $\mathcal{C}(X)\times\mathcal{C}(Y)$ given by $\pi(c_1,c_2)=\pi_u(c_1)\pi_v(c_2)$; that is, the choice functions of the two agents are determined independently.  Let $A\times B$ and $(a,b)\in A\times B$ be arbitrary and observe that $\sum_{c\in\mathcal{C}(X)\times\mathcal{C}(Y)}\pi(c)\mathbf{1}\{c(A,B)=(a,b)\}=\left(\sum_{c_1\in\mathcal{C}(X)}\pi_u(c_1)\mathbf{1}_{c_1(A)=a}\right)\left(\sum_{c_2\in\mathcal{C}(Y)}\pi_v(c_2)\mathbf{1}_{c_2(B)=b)}\right)=\left(\frac{u(a)}{u(A)}\right)\left(\frac{v(b)}{v(B)}\right)$, confirming stochastic separability.  Here, the first equality follows from the definition of $\pi$ and the fact that $\mathbf{1}\{(c_1,c_2)(A,B)=(a,b)\}=\mathbf{1}\{c_1(A)=a\}\mathbf{1}\{c_2(B)=b\}$ and the second from the definition of $\pi_u$ and $\pi_v$.

That any stochastically separable joint choice rule satisfies marginality follows from Theorem~\ref{marginalindependent}.

Finally, suppose that $p$ is a joint Luce rule, with map $f:X\times Y \rightarrow\mathbb{R}_{++}$ and suppose that marginality is satisfied.  Without loss, let us assume that $\sum_{x\in X,y\in Y}f(x,y)=1$.  Define $u(x)=p(x,y|X\times\{y\})$ for any $y\in Y$; by marginality, $u$ is well-defined and $\sum_x u(x)=1$.  Similarly define $v(y)=p(x,y|\{x\}\times Y)$.  Now, for any $x,x'\in X$ and any $y\in Y$, it follows that $\frac{u(x)}{u(x')}=\frac{p(x,y|X\times \{y\})}{p(x',y|X\times\{y\})}=\frac{f(x,y)}{f(x',y)}$.  Consequently, for each $y\in Y$, there exists $\alpha(y)>0$ for which $f(x,y)=\alpha(y)u(x)$.  By marginality, we know that for any $y\in Y$, $v(y)=\sum_{x\in X}p(x,y)=\sum_{x\in X}\alpha(y)u(x)=\alpha(y)$, whereby $\alpha(y)=v(y)$ and consequently $f(x,y)=u(x)v(y)$, as we wanted to show. \end{proof}

\subsection{Theorem \ref{uniqueconvex}}

\begin{definition}
For the marginal graph of a single agent, we call a path $\rho$ a finite sequence of sets $\{A_i\}_{i=0}^{|X|}$ such that $A_{i+1}\subsetneq A_i$ for all $i$, $A_0=X$, and $A_{|X|}=\emptyset$.
\end{definition}

\begin{definition}
For a marginal random choice rule and its corresponding marginal graph, we call a path supported if for all $i \in \{0, \dots, |X|-1\}$, $q(A_i \setminus A_{i+1},A_i)>0$.
\end{definition}

\begin{definition}
We call two paths $\rho$ and $\rho'$ branching if there exists some $i\leq j$ with $i,j \in \{1,\dots,|X|-1\}$ such that $A_{i-1}^{\rho} \neq A_{i-1}^{\rho'}$, $A_{j+1}^{\rho} \neq A_{j+1}^{\rho'}$, and for all $m \in \{i, \dots, j\}$, $A_m^{\rho} = A_m^{\rho'}$.
\end{definition}

\begin{definition}
We call two paths $\rho$ and $\rho'$ in-branching if there exists some $i \in \{1,\dots, |X|-1\}$ such that $A_i^{\rho} = A_i^{\rho'}$ and $A_{i-1}^{\rho} \neq A_{i-1}^{\rho'}$
\end{definition}

\begin{definition}
We call two paths $\rho$ and $\rho'$ out-branching if there exists some $i \in \{1,\dots, |X|-1\}$ such that $A_i^{\rho} = A_i^{\rho'}$ and $A_{i+1}^{\rho} \neq A_{i+1}^{\rho'}$
\end{definition}

\begin{proof}
We begin with a series of observations. Recall that in the marginal graph system a linear order pair is represented by a path along the marginal graph and a common path along each conditional graph of the path along the marginal graph. In the marginal graph system, marginality gives us three things.
\begin{enumerate}
    \item The marginal graph system is well defined as marginal choice probabilities are well defined.
    \item Each conditional graph satisfies inflow equals outflow as marginality is equivalent to recursivity by Lemma 2.
    \item Inflow equals outflow of the marginal graph tells us that all the flow along conditional graphs associated with edges going into a node on the marginal graph must be assigned to conditional graphs of edges leaving that node of the marginal graph. This can also be seen as a direct result of recursivity.
\end{enumerate}

Without loss of generality, let the marginal choices over $X$ be uniquely rationlizable. Consider the marginal graph system with $X$ defining the marginal graph and $Y$ defining the conditional graphs. From \citet{turansick2022identification} we know that each supported path of the marginal graph has an edge that is unique to that supported path. Further, above that edge, for each supported path, the supported path is never in-branching with another supported path. Similarly, below that edge, for each supported path, the supported path is never out-branching with another supported path. These two observations mean that there is only a single supported (sub-)path from $X$ to the chosen edge and a single supported (sub-)path from the chosen edge to a singleton. This means that all the inflow to this edge can be traced back to $X$ and all the outflow from this set can be traced to a singleton. This along with observation (3) above means that for each conditional graph along the considered marginal path, the flows along those conditional graphs must be weakly larger than the flows along the conditional graph of the considered edge of the marginal graph.

Consider the following algorithm.

\begin{enumerate}
    \item Take as input a marginal graph system whose marginal graph is uniquely rationalizable by the random utility model.
    \item Enumerate the set of supported paths on the marginal graph $\{1,\dots,n\}$.
    \item For each supported path $\succ_i$, let $E_i$ denote an/the edge that is unique to that supported path among all supported paths.
    \item Initialize at $i=1$, set $q_1(x,y|A,B)=q(x,y|A,B)$ for all such $(x,y,A,B)$, and $q_1(x,A)=q(x,A)$ for all such $(x,A)$.
    \item For the path $\succ_i$, consider the conditional graph at $E_i$ with edge weight $q(x_i,A_i)$. This conditional graph has non-negative flows (from non-negativity) and satisfies inflow equals outflow (from marginality being equivalent to recursivity (Lemma 2)). This means we can decompose (see the construction in the proof of Lemma 5) the conditional graph into path flows whose total flow sums to $q(x_i,A_i)$. For each path $\succ'_i$ on the conditional graph, set $\nu(\succ_i,\succ'_i)$ equal to the prior stated path flow of $\succ'_i$.
    \item For each conditional graph along the path of $\succ_i$, set $q_{i+1}(w,y|C,B)=q_i(w,y|C,B)-q_i(x_i,y|A_i,B)$. By the logic just prior to the algorithm, this is non-negative. For each conditional graph not along the path of $\succ_i$, set $q_{i+1}(w,y|C,B)=q_i(w,y|C,B)$. For each edge of the marginal graph along the path of $\succ_i$, set $q_{i+1}(w,C)=q_i(w,C)-q_i(x_i,A_i)$. By inflow equals outflow on the marginal graph, this is non-negative. For each edge of the marginal graph not along the path of $\succ_i$, set $q_{i+1}(w,C)=q_i(w,C)$.
    \item Note that the marginal graph system corresponding to iteration $i+1$ satisfies non-negativity and marginality as we are subtracting out a vector that satisfies marginality at every step of the algorithm. If there is any positive flow left on the marginal graph, set $i=i+1$ and return to step (5). If there is no positive flow left on the marginal graph, terminate the algorithm.
\end{enumerate}

As this algorithm terminates with exactly zero flow everywhere along the graph system, this algorithm assigns exactly $q(x,y|A,B)$ to linear order pairs which rank $x$ exactly at the top of $A$ and $y$ exactly at the top of $B$. By Lemma 3, this algorithm has found a rationalization of the random joint choice rule and we are done.

\end{proof}

\bibliographystyle{ecta}
\bibliography{crum}
\end{document}